\documentclass[aps,twocolumn,floats,nofootinbib]{revtex4}
\usepackage {amsmath}
\usepackage{graphics,graphicx,epsfig}
\usepackage{amssymb,color,bm,dcolumn,accents}
\usepackage{epsf,epstopdf,wrapfig}
\usepackage{textgreek}

\newcommand{\beq}{\begin{equation}}
\newcommand{\eeq}{\end{equation}}
\newcommand{\beqn}{\begin{eqnarray}}
\newcommand{\eeqn}{\end{eqnarray}}

\begin{document}

\title{Searching for collective behavior in a small brain} 

\author{Xiaowen Chen,$^1$ Francesco Randi,$^1$ Andrew M. Leifer,$^{1,2}$ and William Bialek$^{1,3,4}$}

\affiliation{$^1$Joseph Henry Laboratories of Physics, $^2$Princeton Neuroscience Institute, and 
$^3$Lewis-Sigler Institute for Integrative Genomics, Princeton University, Princeton, NJ 08544\\
$^4$Initiative for the Theoretical Sciences, The Graduate Center, City University of New York, 365 Fifth Ave., New York, NY 10016}

\date{\today}

\begin{abstract}
In large neuronal networks, it is believed that functions emerge through the collective behavior of many interconnected neurons. Recently, the development of experimental techniques that allow simultaneous recording of calcium concentration from a large fraction of all neurons in \textit{Caenorhabditis elegans}---a nematode with 302 neurons---creates the opportunity to ask if such emergence is universal, reaching down to even the smallest brains. Here, we measure the activity of 50+ neurons in \textit{C.~elegans}, and analyze the data by building the maximum entropy model that matches the mean activity and pairwise correlations among these neurons. To capture the graded nature of the cells' responses, we assign each cell multiple states.  These models, which are equivalent to a family of Potts glasses, successfully predict higher statistical structure in the network.  In addition, these models exhibit signatures of collective behavior: the state of single cells can be predicted from the state of the rest of the network; the network, despite being sparse in a way similar to the structural connectome, distributes its response globally when locally perturbed; the distribution over network states has multiple local maxima, as in models for memory; and the parameters that describe the real network are close to a critical surface in this family of models.
\end{abstract}
                              
\maketitle
\section{Introduction}

The ability of the brain to generate coherent thoughts, percepts, memories, and actions depends on the coordinated activity of large numbers of interacting neurons.  It is an old idea in the physics community that these collective behaviors in neural networks should be describable in the language of statistical mechanics~\cite{hopfield1, hopfield_graded, sompolinsky85}.  For many years it was very difficult to connect these ideas with experiment, but new opportunities are offered by the recent emergence of methods to record, simultaneously, the electrical activity of large numbers of neurons~\cite{dombeck10, zebrafish, retina_data04, nguyen16pnas, nguyen17plos, vivek16pnas}.  In particular, it has been suggested that maximum entropy models \cite{jaynes57} provide a path to construct a statistical mechanics description of network activity directly from real data \cite{schneidman06}, and this approach has been pursued in the analysis of the vertebrate retina as it responds to natural movies and other light conditions~\cite{schneidman06, cocco09retina,gtspinglass,gt14plos},  the dynamics of the hippocampus during exploration of real and virtual environments~\cite{monasson_rosay15,ens17hippocampus, leenoy16}, and  the coding mechanism of spontaneous spikes in cortical networks~\cite{tang08cortical, ohiorhenuan10cortical, koster14cortical}.

Maximum entropy models that match low order features of the data, such as the mean activity of individual neurons and the correlations between pairs, make quantitative predictions about higher order structures in the network, and in some cases these are in surprisingly detailed agreement with experiment~\cite{gt14plos, leenoy16}.  These models also illustrate the collective character of network activity.  In particular, the state of individual neurons often can be predicted with high accuracy from the state of the other neurons in the network, and the models that are inferred from the data are close to critical surfaces in their parameter space, which connects with other ideas about the possible criticality of biological networks~\cite{criticality11, munoz_rmp, leenoy18}.

Thus far, almost all discussion about collective phenomena in networks of neurons has been focused on vertebrate brain, with neurons that generate discrete, stereotyped action potentials or spikes~\cite{spikes}.  This discreteness suggests a natural mapping into an Ising model, which is at the start of the maximum entropy analyses, although one could imagine alternative approaches.  What is not at all clear is whether these approaches could capture the dynamics of networks in which the neurons generate graded electrical responses. An important example of this question is provided by the nematode \textit{Caenorhabditis elegans}, which does not have the molecular machinery needed to generate conventional action potentials~\cite{worm_noap98}. 

\begin{figure*}[t]
\centerline{\includegraphics[width=0.9\textwidth]{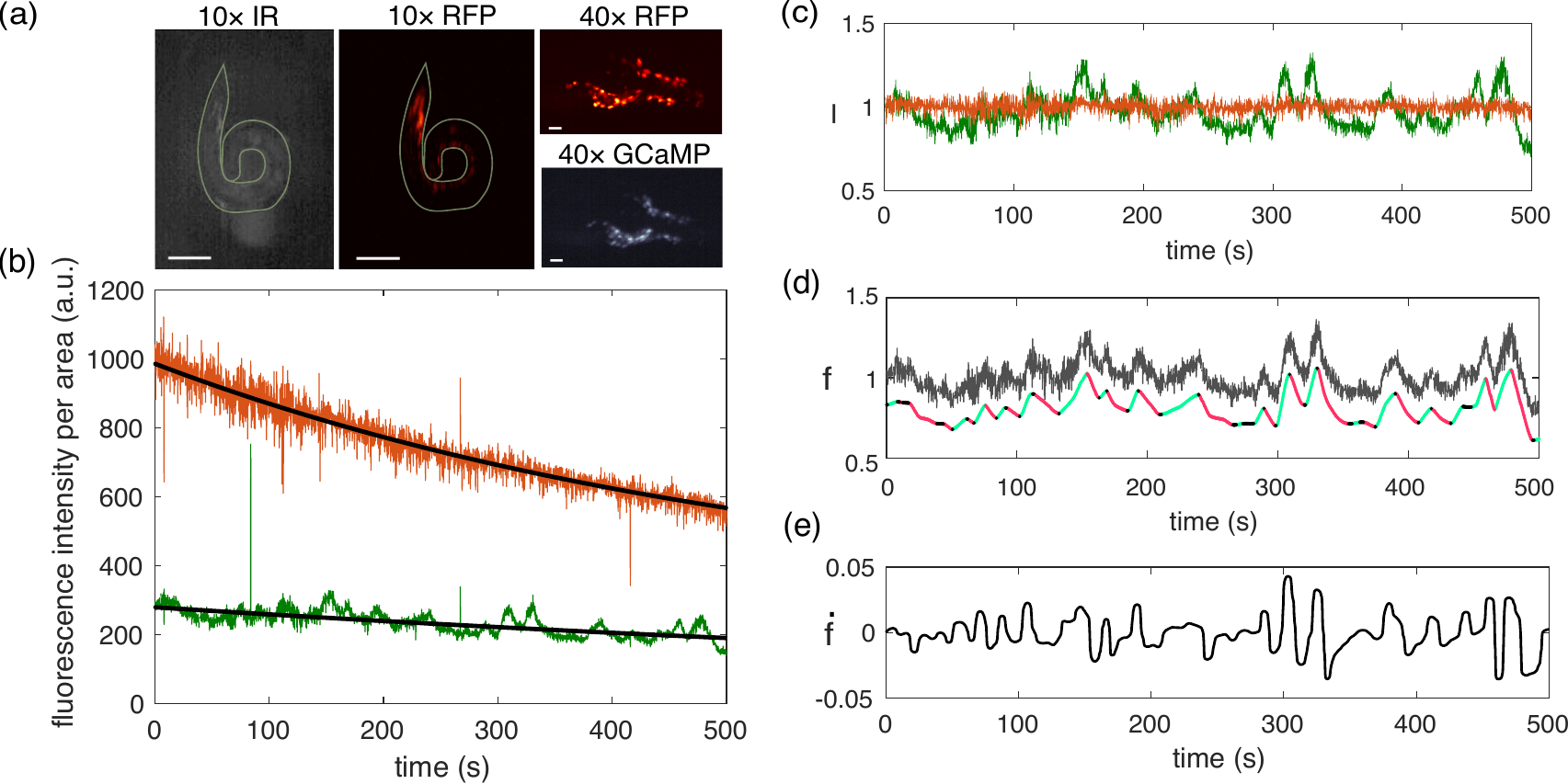}}
\caption{Schematics of data acquisition and processing. (a) Examples of the raw images acquired through the $10\times$ (scale bar equals $100\mu m$) and $40\times$ (scale bar equals $10\mu m$) objectives. The body of the nematode is outlined with light green curves. (b) The intensity of the nuclei-localized fluorescent protein tags---the calcium-sensitive GCaMP and the control fluorophore RFP---are measured as functions of time. Photobleaching occurs on a longer time scale than the intracellular calcium dynamics, which allows us to perform photobleaching correction by dividing the raw signal with its exponential fit, resulting in the signals of panel (c). (d) The normalized ratio of the photobleaching-corrected intensity, $f$, is a proxy for the calcium concentration in each neuron nuclei (dark grey).   As described in the text, this signal is  discretized using the denoised time derivative $\dot{f}$; we use three states, marked as red, blue, and black  after smoothing (lightly offset for ease of visualization). (e) The time derivative $\dot{f}$, extracted using total-variation regularized differentiation.}
\label{fig:set_up_schematics}
\end{figure*}

The nervous system of {\em C.~elegans} has just 302 neurons, yet the worm can still exhibit complex neuronal functions: locomotion, sensing, nonassociative and associative learning, and sleep-wake cycles~\cite{stephens11, celegans_review09, learning_celegans10, sleep17}.  All of the neurons are ``identified,'' meaning that we can find the cell with a particular label in every organism of the species, and in some cases we can find analogous cells in nearby species~\cite{bullock_invert65}. In addition, this is the only organism in which we know the entire pattern of connections among the cells, usually known as the (structural) connectome~\cite{connectome}.  The small size of this nervous system, together with its known connectivity, has always made it a tempting target for theorizing, but relatively little was known about the patterns of electrical activity in the system.  This has changed dramatically with the development of genetically encodable indicator molecules, whose fluorescence is modulated by changes in calcium concentration, a signal which in turn follows electrical activity~\cite{gcamp6}.  Combining these tools with high resolution tracking microscopy opens the possibility of recording the activity in the entire {\em C.~elegans} nervous system as the animal behaves freely~\cite{nguyen16pnas, vivek16pnas,nguyen17plos}.

In this paper we make a first try at the analysis of experiments in {\em C.~elegans} using the maximum entropy methods that have been so successful in other contexts.  Experiments are evolving constantly, and in particular we expect that recording times will increase significantly in the near future.  To give ourselves the best chance of saying something meaningful, we focus on sub--populations of up to fifty neurons, in immobilized worms where signals are most reliable.  We find that, while details differ, the same sorts of models, which match mean activity and pairwise correlations, are successful in describing this very different network. In particular, the models that we learn from the data share topological similarity with the known structural connectome, allow us to predict the activity of individual cells from the state of the rest of the network, and seem to be near a critical surface in their parameter space.

\section{Data acquisition and processing}

Following methods described previously~\cite{nguyen16pnas, nguyen17plos}, nematodes \textit{Caenorhabditis elegans} were genetically engineered to expressed two fluorescent proteins in all of their neurons, with tags that cause them to be localized to the nuclei of these cells.  One of these proteins, GCaMP6s, fluoresces in the green with an intensity that depends on the surrounding calcium concentration, which follows the electrical activity of the cell and in many cases is the proximal signal for transmission across the synapses to other cells~\cite{gcamp6}.  The second protein, RFP, fluoresces in the red and serves as a position indicator of the nuclei as well as a control for changes in the visibility of the nuclei during the course of the experiment.  Parallel control experiments were done on worms engineered to express GFP and RFP, neither of which should be sensitive to electrical activity.  Although our ultimate goal is to understand neural dynamics in the freely moving animal, as a first step we study worms that are immobilized with polystyrene beads, to reduce motion-induced artifacts~\cite{beads}.

As described in Ref.~\cite{nguyen16pnas}, the fluorescence is excited using lasers.  A spinning disk confocal microscope and a high-speed, high-sensitivity Scientific CMOS (sCMOS) camera records red- and green-channel fluorescent image of the head of the worm at a rate of 6 brain-volumes per second at a magnification of $40\times$; a second imaging path records the position and posture of the worm at a magnification of $10\times$, which are used in the tracking of the neurons across different time frames. The raw data thus are essentially movies, and by using a custom machine-learning approach---Neuron Registration Vector Encoding~\cite{nguyen17plos}---we are able to reduce the data to the green and red intensities for each neuron $i$, $I_{i}^g (t)$ and $I_{i}^r (t)$.

As indicated in Fig.~\ref{fig:set_up_schematics}b, the fluorescence intensity undergoes photobleaching, fortunately on much longer time scale than the calcium dynamics. Thus, we can extract the photobleaching effect by modeling the observed fluorescence intensity with an exponential decay:
\begin{equation}
\begin{split}
I_\text{g}(t)&=S_\text{g}(t)(1+\eta_\text{g})(e^{-t/\tau_\text{g}}+A_\text{g})\\
I_\text{r}(t)&=S_\text{r}(t)(1+\eta_\text{r})(e^{-t/\tau_\text{r}}+A_\text{r})
\end{split}
\label{eq:bleach}
\end{equation}
Here, $S_\text{g}(t)$ and $S_\text{r}(t)$ are the true signals corresponding to the calcium concentration, $\eta_\text{g}$ and $\eta_\text{r}$ are stochastic variables representing the noise due to the laser and the camera, $\tau_\text{g}$ and $\tau_\text{r}$ are the characteristic time for photobleaching of the two fluorophores, and $A_\text{g}$ and $A_\text{r}$ represent nonnegative offsets due to a population of unbleachable fluorophores, or regeneration of fluorescent states under continuous illumination.\footnote{One may worry that a constant ``background" fluorescence should be subtracted from the raw signal, rather than contributing to a divisive normalization.  In our data, this background subtraction leads to strongly non-stationary noise in the normalized intensity after the photobleaching correction, in marked contrast to what we find by treating the constant as a contribution from unbleachable or regenerated fluorophores.}

For each neuron, we fit the observed fluorescence intensities to Eqs (\ref{eq:bleach}) with $S_\text{g}(t) = S_\text{g}^0$ and $\eta_\text{g} =0$, and similarly for $S_\text{r}(t)$.  As shown by the black lines in Fig.~\ref{fig:set_up_schematics}b, this captures the slow photobleaching dynamics; we then divide these out to recover normalized intensities in each channel and each cell, $\bar{I}_i^g(t)$ and $\bar{I}_i^r(t)$.  Finally, to reduce instrumental and/or motion induced artifacts, we consider the ratio of the normalized intensities as the signal for each neuron, i.e. $f_i(t) = \bar{I}_i^g(t)/\bar{I}_i^r(t)$ (Fig.~\ref{fig:set_up_schematics}d). In this normalization scheme, if the calcium concentration remains constant, then $f_i(t) = 1$. 

\begin{figure}[t]
\includegraphics[width=\linewidth]{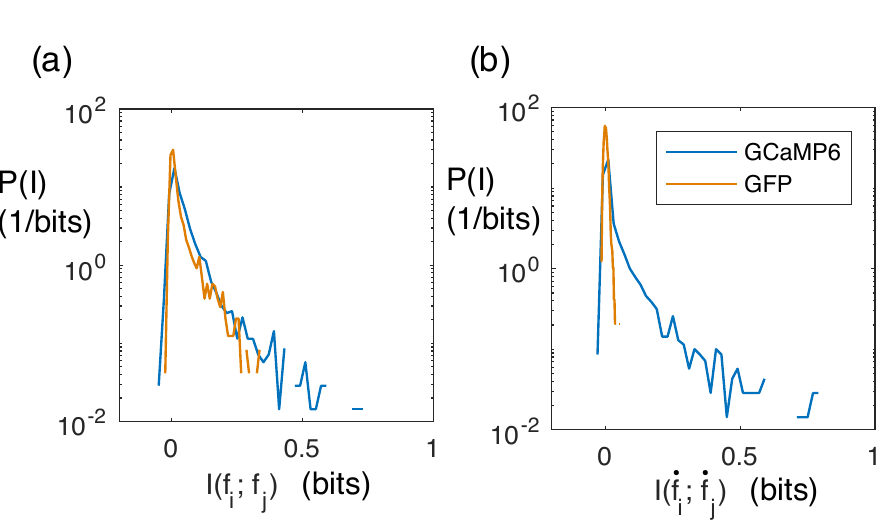}
\caption{Comparison of pairwise mutual information distribution for the calcium-sensitive GCaMP worms and the GFP control worms. Mutual information is estimated using binning and finite-sample extrapolation methods as described in~\citep{mi05} for all pairs of neurons. For the normalized fluorescence ratio, $f$, the distribution of the mutual information, $P(I(f_i; f_j))$, exhibits little difference between the calcium-sensitive GCaMP worm and the GFP control worm (panel (a)). In comparison, for the time derivative of the normalized fluorescence ratio, $\dot{f}$, the distribution of the mutual information, $P(I(\dot{f_i}; \dot{f_j}))$, is peaked around zero for the GFP control worm, while the distribution is wide for the calcium-sensitive GCaMP worm (panel (b)). This observation suggests that time derivative of fluorescence ratio, $\dot{f_i}$, is more informative than its magnitude, $f_i$.}
\label{fig:mi}
\end{figure}

Our goal is to write a model for the joint probability distribution of activity in all of the cells in the network.  To stay as close as possible to previous work, at least in this first try, it makes sense to quantize the activity into discrete states.  One possibility is to discretize based on the magnitude of the fluorescence ratio $f_i(t)$.  But this is problematic, since even in ``control'' worms where the fluorescence signal should not reflect electrical activity, variations in different cells are correlated; this is illustrated in Fig.~\ref{fig:mi}a, where we see that the distribution of mutual information between $f_i(t)$ and $f_j(t)$, across all pairs $(i, j)$, is almost the same in control and experimental worms. A closer look at the raw signal suggests that normalizing by the RFP intensity is not enough  to correct for occasional wobbles of the worm; this causes  the distribution of the fluorescence ratio  to be non-stationary, and generates spurious correlations.   This suggests that (instantaneous) fluorescence signals are not especially reliable, at least given the current processing methods and the state of our experiments. An alternative is to look at the derivatives of these signals, which by definition suffer from the global noise only at a few instances; now there is very little mutual information between $\dot f_i(t)$ and $\dot f_j(t)$ in the control worms, and certainly much less than in the experimental worms, as seen in Fig.~\ref{fig:mi}b.

\begin{figure*}[t]
\centerline{\includegraphics[width=0.8\linewidth]{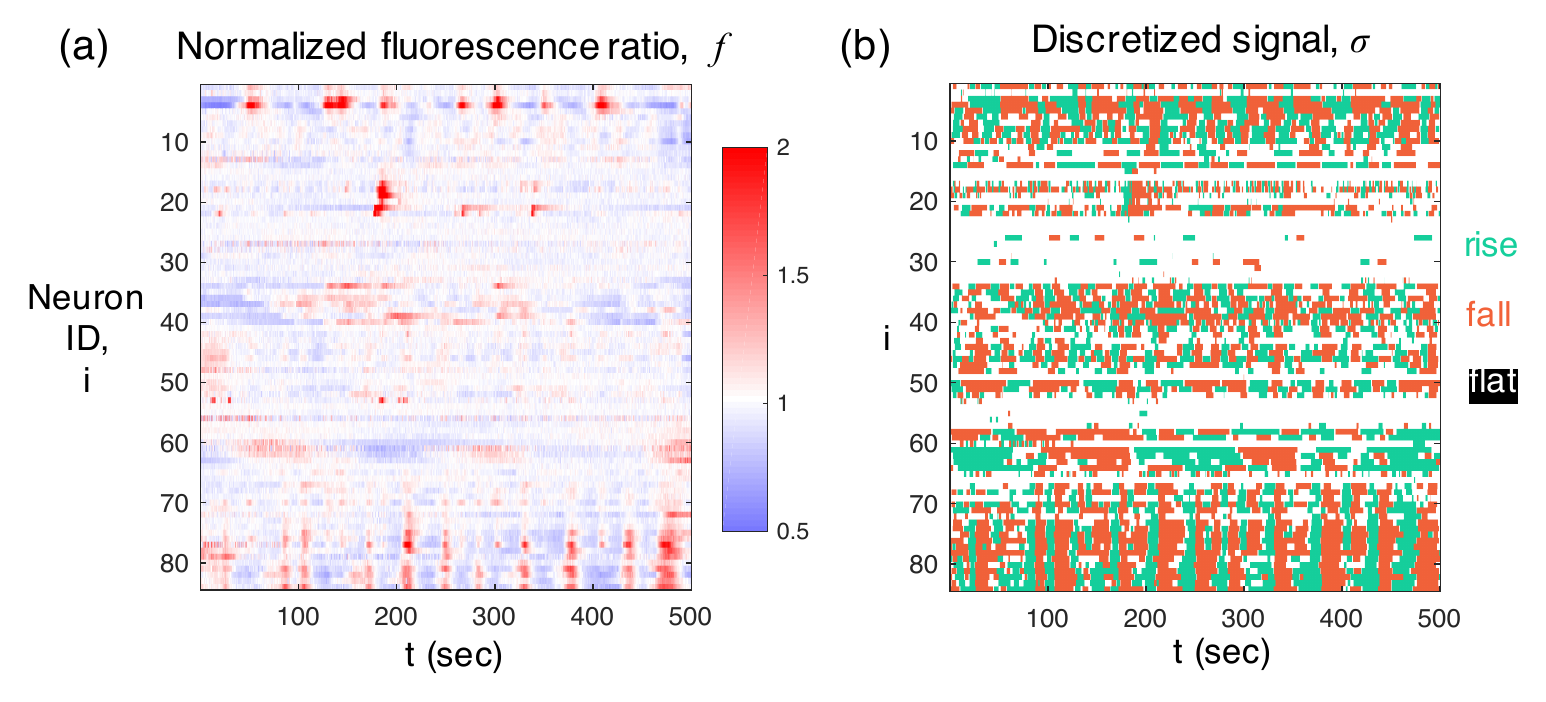}}
\caption{Discretization of the empirically observed fluorescence signals. (a) Heatmap of the normalized fluorescence ratio between photobleaching-corrected GCaMP fluorescence intensity and RFP fluorescence intensity, $f$, for each neuron as a function of time.  (b) Heatmap of the neuronal activity after discretization based on time derivatives of $f$. Green corresponds to a state of ``rising'', red ``falling'', and white ``flat''. }
\label{fig:data_heatmap}
\end{figure*}

To give ourselves a bit more help in isolating a meaningful signal, we denoise the time derivatives.  The optimal Bayesian reconstruction of the underlying time derivative signal $u(t)$ combines a description of noise in the raw fluorescence signal $f(t)$ with some prior expectations about the signal $u$ itself.  We approximate the noise in $f$ as Gaussian and white, which is consistent with what we see at high frequencies, and we assume that the temporal variations in the derivative are exponentially distributed and only weakly correlated in time.  Then maximum likelihood reconstruction is equivalent to minimizing
\begin{equation}\label{eq:tvreg}
\begin{split}
F(u) & = \frac{\tau_f}{\sigma_f}\int_0^{T}dt \lvert \dot{u}\rvert+\frac{1}{2\sigma_n^2 \tau_n}\int_0^{T}dt\lvert Au-f\rvert^2\,\mbox{,}
\end{split}
\end{equation}
where $A$ is the antiderivative operator, the combination $\sigma_n^2\tau_n$ is the spectral density of noise floor that we see in $f$ at high frequencies, while $\sigma_f$ is the total standard deviation of the signal and $\tau_f$ is the typical time scale of these variations; for more on these reconstruction methods see Ref.~\cite{total_var11}.  We determine the one unknown parameter $\tau_f$ by asking that, after smoothing, the cumulative power spectrum of the residue $Au - f$ has the least root mean square difference from the cumulative power spectrum of the extrapolated white noise. 

As an example, Fig.~\ref{fig:set_up_schematics}e shows the smooth derivative of the trace in Fig.~\ref{fig:set_up_schematics}d. After the smooth derivative $u$ is estimated, we discretized the smooth estimate of the signal, $Au$, into three states of ``rise,'' ``fall,'' and ``flat,'' depending on whether the derivative $u$ exceeds a constant multiple of $\sigma_n/\tau_f$, the expected standard deviation of the smooth derivative extracted from a pure white noise. The constant is chosen to be $\sigma_n/\tau_f = 5$, such that the GFP control worm has almost all pairwise mutual information being zero after going through the same data processing pipeline.  An example of the raw fluorescence and final discretized signals is shown in Fig.~\ref{fig:data_heatmap}.

\section{Maximum entropy model}

After preprocessing, the state of each neuron is described by a Potts variable $\sigma_i$, and the state of the entire network is $\{\sigma_i\}$.   As in previous work on a wide range of biological systems~\cite{schneidman06, gt14plos, leenoy16, weigt08, antibody10, birdflock1}, we use a maximum entropy approach to generate relatively simple approximations to the distribution of states, $P(\{\sigma_i\})$, and then ask how accurate these models are in making predictions about higher order structure in the network activity.

The maximum entropy approach begins by choosing some set of observables, $\mathcal{O}_\mu (\{\sigma_i\})$, over the states of the system, and we insist that any model we write down for $P(\{\sigma_i\})$ match the expectation values for these observables that we find in the data,
\begin{equation}
\sum_{\{\sigma_i\}} P(\{\sigma_i\})\mathcal{O}_\mu (\{\sigma_i\}) = \langle \mathcal{O}_\mu (\{\sigma_i\})\rangle_{\rm expt}.
\label{eq:match}
\end{equation}
Among the infinitely many distributions consistent with these constraints, we choose the one that has the largest possible entropy, and hence no structure beyond what is needed to satisfy the constraints in Eq.~(\ref{eq:match}).  The formal solution to this problem is
\begin{equation}
P(\{\sigma_i\}) = \frac{1}{Z} \exp\left[ {-\sum_{\mu}\lambda_\mu \mathcal{O}_\mu (\{\sigma_i\})}\right],
\label{eq:me1}
\end{equation}
where coupling constant $\lambda_\mu$ must be set to satisfy and Eq.~(\ref{eq:match}), and the partition function $Z$ as usual enforces normalization.

Following the original application of maximum entropy methods to neural activity~\cite{schneidman06}, we choose as observables the mean activity of each cell, and the correlations between pairs of cells.  With neural activity described by three states, ``correlations'' could mean a whole matrix or tensor of joint probabilities for two cells to be in particular states.  We will see that models which match this tensor have too many parameters to be inferred reliably from the data sets we have available, and so we take a simpler view in which ``correlation'' measures the probability that two neurons are in the same state.  Equation (\ref{eq:me1}) then becomes
\begin{equation}\label{eq:boltzmann}
P(\sigma) =  \frac{1}{Z}e^{-\mathcal{H(\sigma)}}\,\mbox{,}
\end{equation}
with the effective Hamiltonian
\begin{equation}\label{eq:hamiltonian}
\mathcal{H}(\sigma) =  -\frac{1}{2}\sum_{i\neq j}J_{ij}\delta_{\sigma_i\sigma_j} - \sum_{i}\sum_{r=1}^{p-1} h_i^r\delta_{\sigma_i r}\,\mbox{.}
\end{equation}
The number of states $p = 3$, corresponding to ``rise,'' ``fall,'' and ``flat'' as defined above. The parameters are the pairwise interaction $J_{ij}$ and the local fields $h_i^r$, and these must be set to match the experimental values of the correlations
\begin{equation}\label{eq:cij_def}
c_{ij} \equiv \langle \delta_{\sigma_i\sigma_j} \rangle  = \frac{1}{T}\sum_{t=1}^T \delta_{\sigma_i(t)\sigma_j(t)} \,\mbox{,}
\end{equation}
and the magnetizations
\begin{equation}\label{eq:mir_def}
m_i^r \equiv \langle \delta_{\sigma_i r}  \rangle =  \frac{1}{T}\sum_{t = 1}^T \delta_{\sigma_i(t) r} \,\mbox{.}
\end{equation}
Note that the local field for the ``flat" state, $h_i^p$, is set to zero by convention. In addition, the interaction $J_{ij}$ can be non-zero for any pairs of neurons $i$ and $j$ regardless of the positions of the neurons (both physical and in the structural connectome), i.e. the equivalent Potts model does not have a pre-defined spatial structure.

\begin{figure}[t]
\centering
\includegraphics[width=\linewidth]{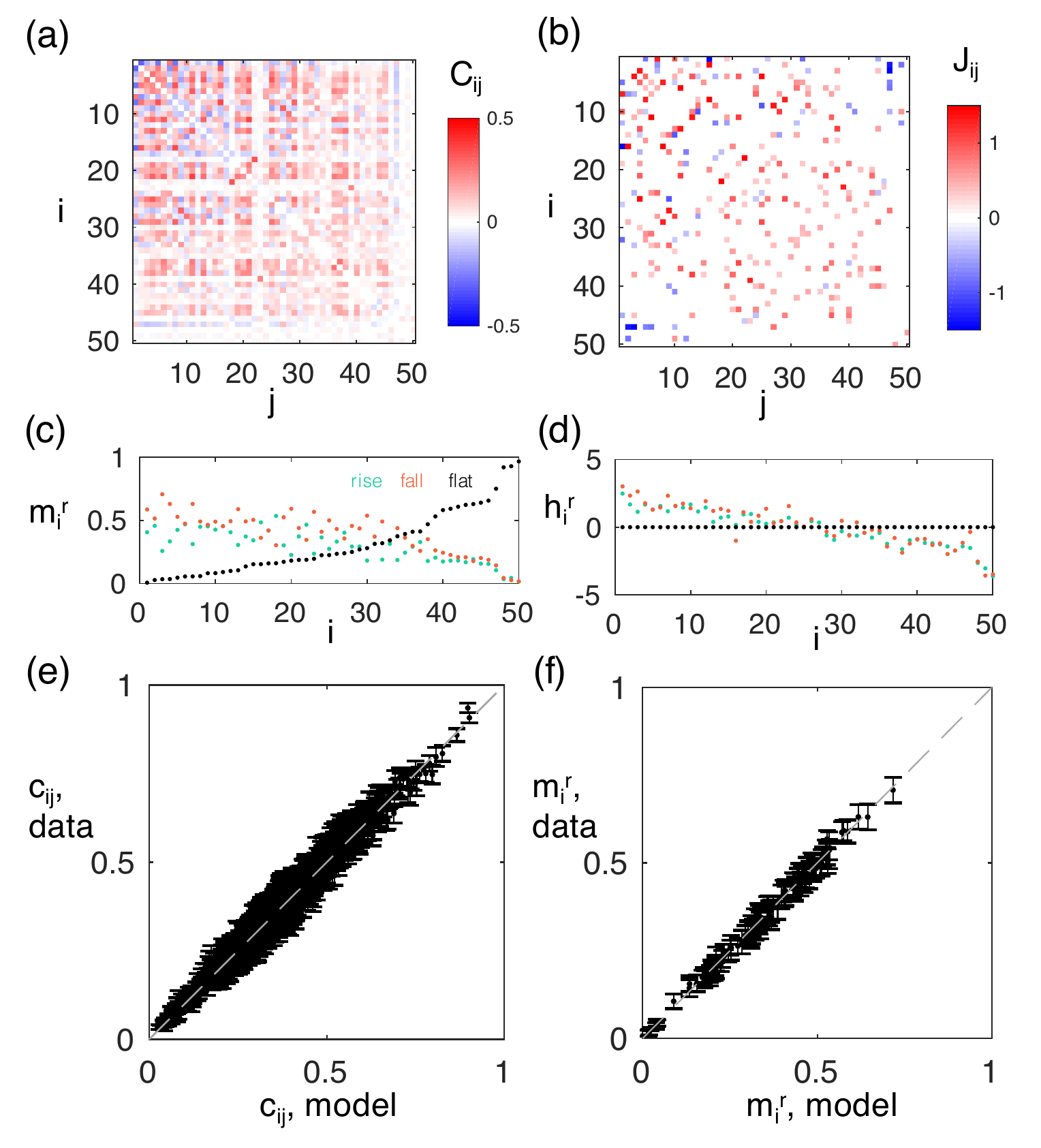}
\caption{\textit{Model construction:} learning the maximum entropy model from data. (a) Connected pairwise correlation matrix, $C_{ij}$, measured for a subgroup of 50 neurons. (b) The inferred interaction matrix, $J_{ij}$. (c) Probability of neuron $i$ in state $r$, for the same group of 50 neurons as panel (a). (d) The inferred local field, $h_i^r$. (e) Model reproduces pairwise correlation (unconnected) within variation throughout the experiment. Error bars are extrapolated from bootstrapping random halves of the data. (f) Same as panel (e), but for mean neuron activity $m_i^r$.} 
\label{fig:model_construction}
\end{figure} 

The model parameters are learned using coordinate descent and Markov chain Monte Carlo (MCMC) sampling~\cite{dudik04, broderick2007faster, ugm}. In particular, we initialize all parameters at zero. For each optimization step, we calculate the model prediction $c_{ij}$ and $m_i^r$ by alternating between MCMC sampling with $10^4$ MC sweeps and histogram sampling to speed up the estimation. Then, we choose a single parameter from the set of parameters $\lbrace J_{ij}, h_i^r \rbrace$ to update, such that the increase of likelihood of the data is maximized~\cite{dudik04}. We repeat the observable estimation and parameter update steps until the model reproduces the constraints within the experimental errors, which we estimate from variations across  random halves of the data. This training procedure leaves part of the interaction matrix $J_{ij}$ zero, while the model is able to reproduce the magnetization $m_i^r$ and the pairwise correlation $c_{ij}$ within the experimental errors (Fig.~\ref{fig:model_construction}). 

Because of the large temporal correlation in the data, the number of independent data in the recording is small compared to the number of parameters. This makes us worry about overfitting, which we test by randomly selecting $5/6$ of the data as training set, inferring the maximum entropy model from this training set, and then comparing the log-likelihood of both the training data and the test data with respect to the maximum entropy model. No signs of overfitting are found for subgroups of up to $N = 50$ neurons, as indicated by that fact that the difference of the log-likelihood is zero within error bars (Fig.~\ref{fig:overfit}; details in Appendix A).   This is not true if we try to match the full tensor correlations (Appendix B), which is why we restrict ourselves to the simpler model.

\begin{figure}
\includegraphics[width=\linewidth]{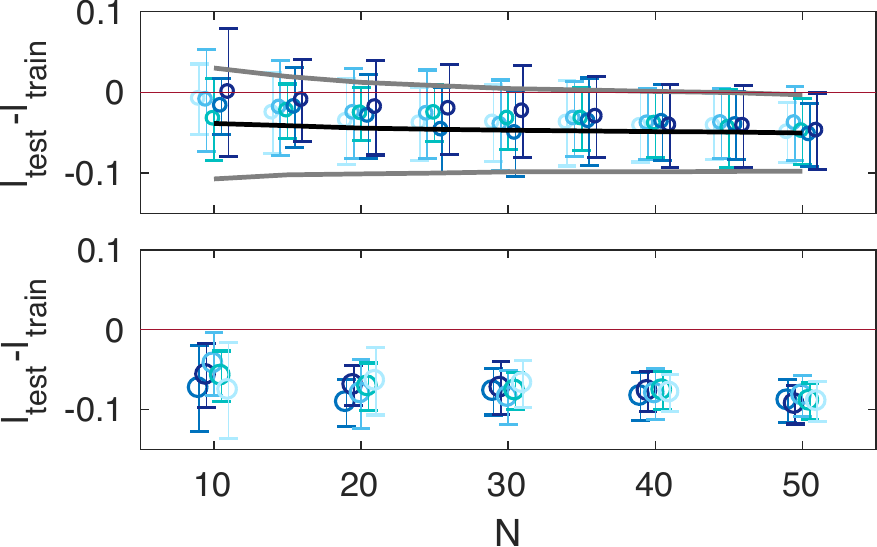}
\caption{\textit{Top}:  No signs of overfitting are observed for models of up to $N = 50$ neurons, measured by the difference of per-neuron log-likelihood of the data under the pairwise maximum entropy model for training sets consists of $5/6$ of the data and test sets. Clusters around $N = 10, 15, 20, \dots, 50$ represent randomly chosen subgroups of $N$ neurons. Error bars are the standard deviation across 10 random partitions of training and test samples. The dashed lines show the expected per-neuron log-likelihood difference and its standard deviation calculated through perturbation methods (see Appendix A).  \textit{Bottom}:  The difference between log likelihood of the training data and of the test data is greater than 0 (the red line) within error bars for maximum entropy models on $N = 10, 20, \dots, 50$ neurons with pairwise correlation tensor constraint (see Appendix B), which suggests that this model does not generalize well.} 
\label{fig:overfit}
\end{figure}

\section{Does the model work?}

The maximum entropy model has many appealing features, not least its mathematical equivalence to statistical physics problems for which we have some intuition.  But this does not mean that this model gives an accurate description of the real network.  Here we test several predictions of the model.  In practice we generate these predictions by running a long Monte Carlo simulation of the model, and then treating the samples in this simulation exactly as we do the real data.  We emphasize that, having matched the mean activity and pairwise correlations, there are no free parameters, so that everything which follows is a prediction and not a fit.

\begin{figure}
\includegraphics[width=\linewidth]{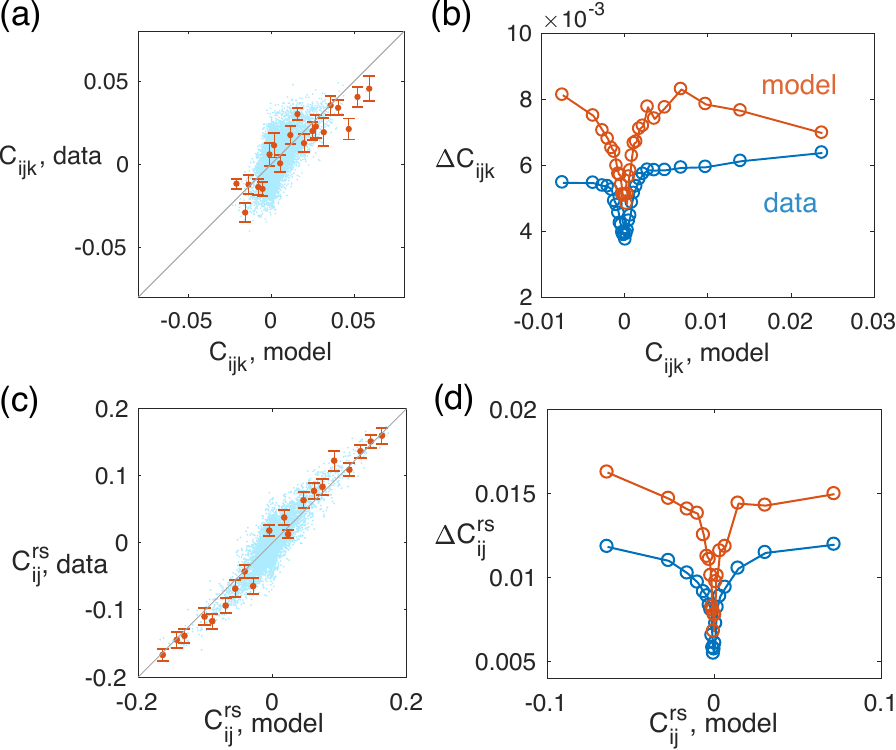}
\caption{\textit{Model validation:} The model predicts unconstrained higher order correlations of the data. Panel (a) shows the comparison between model prediction and data for the connected three-point correlation $C_{ijk}$ for a representative group of $N = 50$ neurons. All 19800 possible triplets are plotted with the blue dot. Error bars are generated by bootstrapping random halves of the data, and are shown for 20 uniformly spaced random triplets in red. Panel (b) shows the error of three-point function $\Delta C_{ijk}$ as a function of the connected three-point function $C_{ijk}$, binned by its value predicted by the model, $C_{ijk}, model$. The red curve is the difference between data and model prediction. The blue curve is the standard error from mean of $C_{ijk}$ over the course of the experiment, extracted by bootstrapping random halves of the experiment. Panels (c, d) are the same as panels (a, b), but for the connected two-point correlation tensor $C_{ij}^{rs}$.}
\label{fig:higher_order_correlation_check}
\end{figure}

Since we use the correlations between pairs of neurons in constructing our model, the first nontrivial test is to predict correlations among triplets of neurons,
\begin{equation}
C_{ijk}= \sum_{r=1}^p \langle(\delta_{\sigma_i r}-\langle  \delta_{\sigma_i r} \rangle)
(\delta_{\sigma_j r}-\langle \delta_{\sigma_j r}\rangle)
(\delta_{\sigma_k r}-\langle \delta_{\sigma_k r}\rangle)\rangle\,\mbox{.}
\end{equation}
More subtly, since we used only the probability of two neurons being in the same state, we can try to predict the full matrix of pairwise correlations,
\begin{equation}
C_{ij}^{rs}\equiv \langle \delta_{\sigma_i r}\delta_{\sigma_j s}\rangle - \langle \delta_{\sigma_i r}\rangle \langle \delta_{\sigma_j s}\rangle\,\mbox{;}
\end{equation}
note that the trace of this matrix is what we used in building the model.  Scatter plots of observed vs predicted values for 
$C_{ijk}$ and $C_{ij}^{rs}$ are shown in Fig.~\ref{fig:higher_order_correlation_check}a and c.  In parts b and d of that figure we pool the data, comparing the root-mean-square differences between our predictions and mean observations (model error) with errors in the measurements themselves.  Although not perfect, model errors are always within $1.5\times$ the measurement errors, over the full dynamic range of our predictions.

\begin{figure}[b]
\centering
\includegraphics[width=\linewidth]{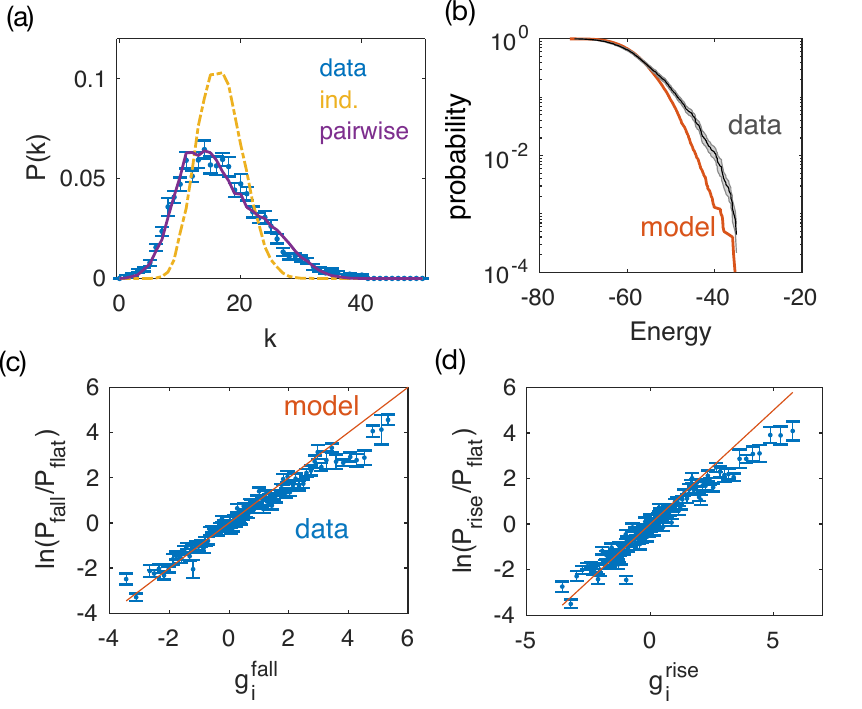}
\caption{\textit{Model validation:} comparison between model prediction and data for observables not constrained by the model. The neuron network has $N = 50$ neurons. (a) Probability of $k$ neurons being in the same state. Blue dots are computed from the data. Yellow dash-dot line is the prediction from a model where all neurons are independent, generated by applying a random temporal cyclic permutation to the activity of each neuron. Purple line is the prediction of the pairwise maximum entropy model. (b) Tail distribution of the energy for the data and the model. All error bars in this figure are extrapolated from bootstrapping. (c, d) Probability ratio of the state of a single neuron as a function of the effective field $g_i^r$, binned by the value of the effective field. Error bars are the standard deviation after binning.}
\label{fig:model_data_check}
\end{figure}

Turning to more global properties of the system, we consider the probability of $k$ neurons being in the same state, defined as
\begin{equation}
P(k) \equiv \Big \langle \sum_{r=1}^p I_{\sum_{i=1}^N\delta_{\sigma_i r}=k} \Big\rangle\,\mbox{,}
\end{equation} 
where $I$ is the indicator function. It is useful to compute this distribution not just from the data, but also from synthetic data in which we break correlations among neurons by shifting each cell's sequence of states by an independent random time.  We see in Fig.~\ref{fig:model_data_check}a that the real distribution is very different from what we would see with independent neurons, so that in particular the tails provide a signature of correlations.  These data agree very well with the distributions predicted by the model.

Our model assigns an ``energy'' to every possible state of the network [Eq.~(\ref{eq:hamiltonian})], which sets the probability of that state according to the Boltzmann distribution.  Because our samples are limited, we cannot test whether the energies of individual states are correct, but we can ask whether the distribution of these assigned energies across the real states taken on by the network agree with what it predicted from the model.  Figure \ref{fig:model_data_check}b compares these distributions, shown cumulatively, and we see that there is very good overlap between theory and experiment across $\sim 90\%$ of the density, with the data having a slightly fatter tail than predicted.  The good agreement extends over a range of $\Delta E\sim 20$ in energy, corresponding to predicted probabilities that range over a factor of $\exp(\Delta E) \sim 10^8$.

The maximum entropy model gives the probability for the entire network to be in a given state, which means that we can also compute the conditional probabilities for the state of one neuron given the state of all the other neurons in the network. Testing whether we get this right seems a very direct test of the idea that activity in the network is collective.  This conditional probability can be written as 
\begin{equation}
P(\sigma_i | \{\sigma_{j\neq i}\}) \propto \exp\left[ \sum_{r=1}^{p-1} g_i^r \delta_{\sigma_i r}\right] ,
\end{equation}
where the effective fields are combinations of the local field $h_i^r$ and each cell's interaction with the rest of the network.
\begin{equation}
g_i^r = h_i^r +\sum_{ j\neq i}^N J_{ij}(\delta_{\sigma_jr}-\delta_{\sigma_jp}) \,\mbox{.}
\end{equation}
Then the probabilities for the states of neuron $i$ are set by
\begin{equation}
\frac{P(\sigma_i = r)}{P(\sigma_i = p)} = e^{g_i^r}\,\mbox{,}
\label{eq:single_neuron1}
\end{equation} 
where the last state $p$ is a reference.  In Figure~\ref{fig:model_data_check}c and d we test these predictions.  In practice we walk through the data, and at each moment in time, for each cell, we compute effective fields.  We then find all moments where the effective field falls into a small bin, and compute the ratio of probabilities for the states of the one cell, collecting the data as shown.  The agreement is excellent, except at extreme values of the field which are sampled only very rarely in the data. We note the agreement extends over a dynamic range of roughly two decades in the probability ratios.\footnote{The claim that behaviors are collective requires a bit more than predictability.  It is possible that behaviors of individual cells are predictable from the state of the rest of the network, but that most of the predictive power comes from interaction with a single strongly coupled partner.  We have checked that the mutual information $I(\sigma_i ; g_i^r)$ is larger than the maximum of $I(\sigma_i ; \sigma_k)$, in almost all cases.}

\section{What does the model teach us?}

\subsection{Energy landscape}

\begin{figure*}
\centerline{\includegraphics[width=0.9\linewidth]{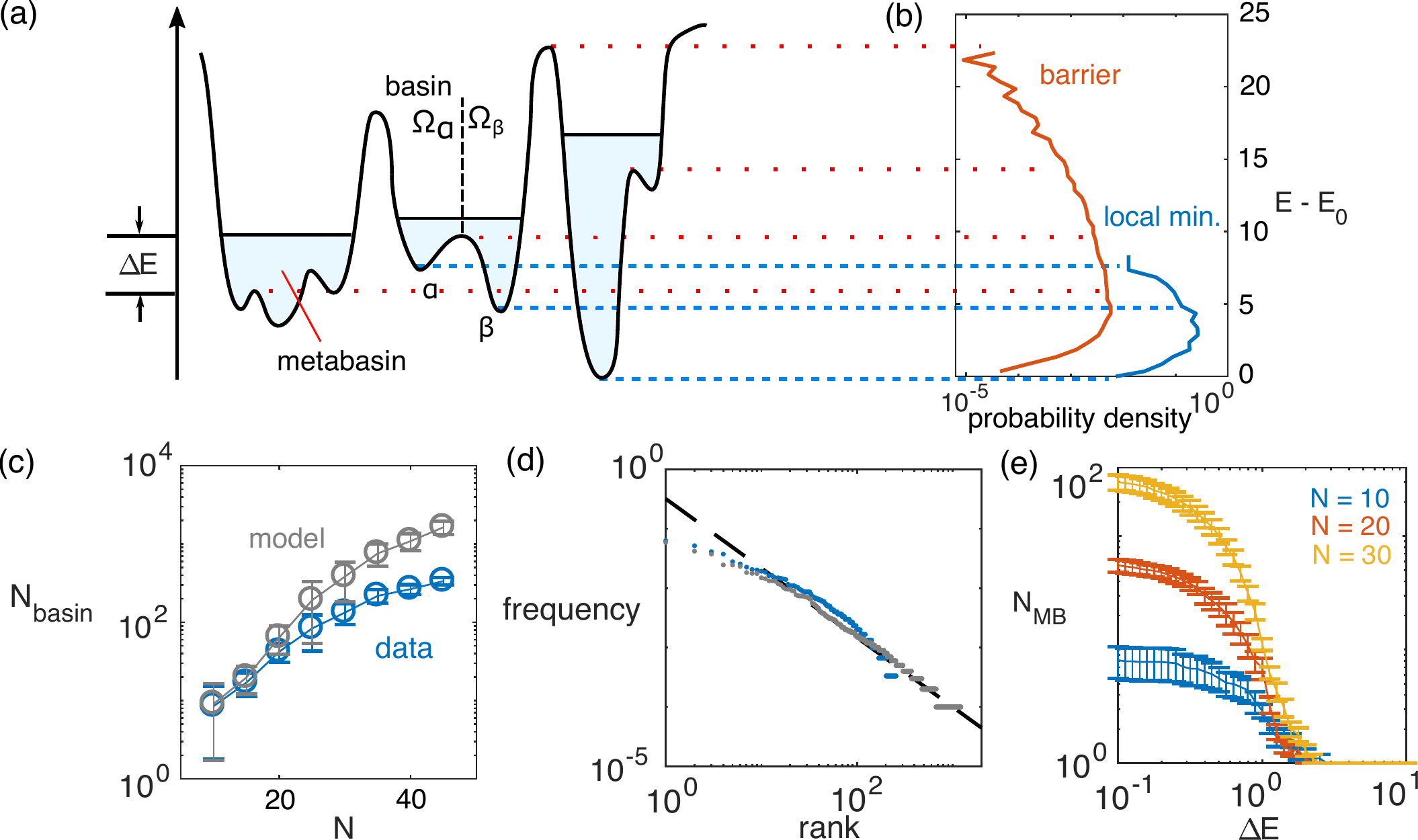}}
\caption{Energy landscape of the inferred maximum entropy model. (a) Schematic of the energy landscape with local minima $\alpha$, $\beta$ and the corresponding basin $\Omega_\alpha$, $\Omega_\beta$. Colored in light blue is the metabasin formed at the given energy threshold, $\Delta E$. (b) Typical distribution of the value of the energy minima and the barriers of a maximum entropy model on $N = 30$ neurons. The global energy minimum, $E_0$, is subtracted from the energy, $E$. (c) The number of energy minima increases sub-exponentially as number of neurons included in the model increases. Error bars are the standard deviation of 10 different subgroups of $N$ neurons. (d) The rank-frequency plot for frequency of visiting each basin matches well between data and model for a typical subgroup of 40 neurons.  (e) The number of metabasins, grouped according to the energy barrier, diverges when the energy threshold $\Delta E$ approaches 1 from above.}
\label{fig:energy_landscape}
\end{figure*}

Maximum entropy models are equivalent to Boltzmann distributions and thus define an energy landscape over the states of the system, as shown schematically in Fig.~\ref{fig:energy_landscape}a.  In our case, as in other neural systems, the relevant models have interactions with varying signs, allowing the development of frustration and hence a landscape with multiple local minima.  These local minima are states of high probability, and serve to divide the large space of possible states into basins.  It is natural to ask how many of these basins are supported in subnetworks of different sizes.

To search for energy minima, we performed quenches from initial conditions corresponding to the states observed in the experiment, as described in~\cite{gt14plos}. Briefly, at each update, we change the state of one neuron such that the decrease of energy is maximized, and we terminate this procedure when no single spin flip will decrease the energy; the states that are attracted to local energy minimum $\alpha$ form a basin of attraction $\Omega_\alpha$.  As shown in Fig.~\ref{fig:energy_landscape}c, the number of energy minima grows sub-exponentially as the number of neurons increases.  Note that this approach only gives us the states that the animal has access to, rather than all metastable states, whose number is approximated by greedy quench along a long MCMC trajectory. Nonetheless, the probability of visiting a basin is similar between the data and the model, shown by the rank-frequency plot (Fig.~\ref{fig:energy_landscape}d).

Whether the energy minima correspond to well defined collective states depends on the heights of the barriers between states. Here, we calculate the barrier height between basins by single-spin-flip MCMC, initialized at one minimum $\alpha$ and terminating when the state of the system belongs to a different basin $\Omega_\beta$; the barrier between basins $\Omega_\alpha$ and $\Omega_\beta$ is defined as the maximum energy along this trajectory. This sampling procedure is repeated 1000 times for each initial basin to compute the mean energy barrier. As shown in Fig.~\ref{fig:energy_landscape}b, the distribution of barrier energies strongly overlaps the distribution of the energy minima, which implies that the minima are not well separated.  

Further visualization of the topography of the energy landscape is performed by constructing metabasins, following Ref~\cite{disconnectivity_graph}. Here, we construct metabasins by grouping the energy minima according to the barrier height; basins with barrier height lower than a given energy threshold, $\Delta E$, are grouped into a single metabasin. This threshold can be varied: at high enough threshold, the system effectively does not see any local minima; at low threshold, the partition of the energy landscape approaches the partition given by the original basins of attraction.   If the dynamics were just Brownian motion on the landscape, states within the same metabasin would transition into one other more rapidly than states belonging to different metabasins.  As shown in Fig.~\ref{fig:energy_landscape}e, there is a transition at $\Delta E \approx 1.2$ from single to multiple metabasins for all $N = 10, 20,$ and $30$.  Since the dynamics of the real system do not correspond to a simple walk on the energy landscape (Appendix C and Fig.~\ref{fig:dynamics}), we cannot conclude that this is a true dynamical transition, but it does suggest that the state space is organized in ways that are similar to what is seen in systems with such transitions.

\subsection{Criticality} 

Maximum entropy models define probability distributions that are equivalent to equilibrium statistical physics problems.  As these systems become large, we know that the parameter space separates into distinct phases, separated by critical surfaces.  In several biological systems that have been analyzed there are signs that these critical surfaces are not far from the operating points of the real networks, although the interpretation of this result remains controversial ~\cite{criticality11}.  Here we ask simply whether the same pattern emerges in {\em C.~elegans}.

\begin{figure}
\includegraphics[width=\linewidth]{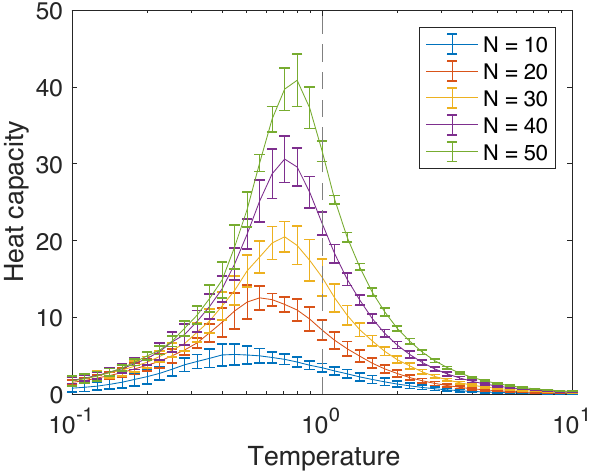}
\caption{The heat capacity is plotted against temperature for models with different number of neurons, $N$. The maximum of the heat capacity approaches the operational temperature of the \textit{C. elegans} neural system $T_0 = 1$ from below as $N$ increases. Error bars are the standard error across 10 random subgroups of $N$ neurons.}
\label{fig:heat_capacity}
\end{figure}

One natural slice through the parameter space of models corresponds to changing the effective temperature of the system, effectively scaling all terms in the log probability up and down uniformly. Concretely, we replace $\mathcal{H(\sigma)}\rightarrow \mathcal{H(\sigma)}/T$ in Eq~(\ref{eq:boltzmann}).  We monitor the heat capacity of the system, as we would in thermodynamics; here the natural interpretation is of the heat capacity as being proportional to the variance of the log probability, so it measures the dynamic range of probabilities that can be represented by the network.  Results are shown in Fig.~\ref{fig:heat_capacity}, for randomly chosen subsets of $N=10, 20, ... , 50$ neurons.  A peak in heat capacity often signals a critical point, and here we see that the maximum of the heat capacity approaches the operational temperature $T_0 = 1$ from below as $N$ becomes larger, suggesting that the full network is near to criticality.

\subsection{Network topology}

\begin{figure}[b]
\centering
\includegraphics[width=\linewidth]{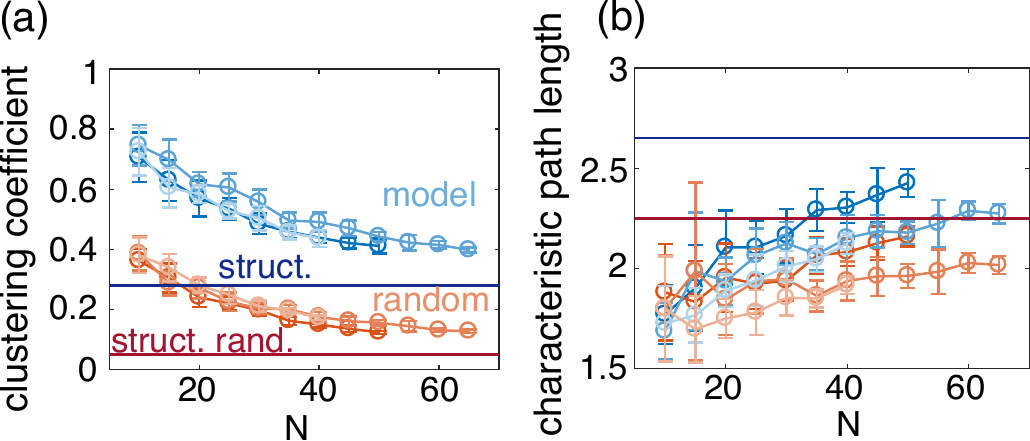}
\caption{The topology of the learned maximum entropy model approaches that of the structural connectome, as the number of neurons being modeled, $N$, increases. The two global topological properties being measured are the clustering coefficient $C$ (panel (a)) and characteristic path length $L$ (panel (b)). Here, the inferred network topology for three different worms is plotted in blue. Red curves are for the randomized network with the same number of neurons, $N$, and number of connections, $N_E$, as the model, where we expect $L_\text{random} \sim \ln(N)/\ln(2N_E/N)$ and $C_\text{random} \sim 2N_E/N^2$. The dark blue line corresponds to the network property of the structural connectome; the dark red line corresponds to randomized network with number of nodes and edges equal to those of the structural connectome~\cite{small_world}. Error bars are generated from the standard deviation across different 10 subgroups of $N$ neurons.}
\label{fig:topology}
\end{figure}

The worm {\em C.~elegans} is special in part because it is the only organism in which we know (essentially) the full pattern of connectivity among neurons.  Our models also have a ``connectome,'' since only a small fraction of the possible pairs of neurons are linked by a nonzero value of $J_{ij}$.  The current state of our experiments is such that we cannot identify the individual neurons, and so we cannot check if the effective connectivity in our model is similar to the anatomical connections.  But we can ask statistical questions about the connections, and we focus on two global properties of the network: the clustering coefficient $C$, defined as the fraction of actual links compared to all possible links connecting the neighbors of a given neuron, averaged over all neurons; and the characteristic path length $L$, defined as the average shortest distance between any pair of neurons. As shown in Fig.~\ref{fig:topology}, the topology of the inferred networks for all three worms that we investigated  differ from random Erd\H os-R\'enyi  graphs with the same number of nodes (neurons) and links (non-zero interactions). Moreover, as we increase the number of neurons that we consider, the clustering coefficient $C$ and the characteristic path length $L$ approaches that found in the structural connectome~\cite{small_world}.

\subsection{Local perturbation leads to global response}

How well can the sparsity of the inferred network explain the observed globally-distributed pairwise correlation? In particular, we would like to examine the response of the network to local perturbations. This test is of particular interest, since its predictions can be examined experimentally, as local perturbation of the neural network can be achieved through optogenetic clamping or ablation of individual neurons.

\begin{figure}[b]
\includegraphics[width=\linewidth]{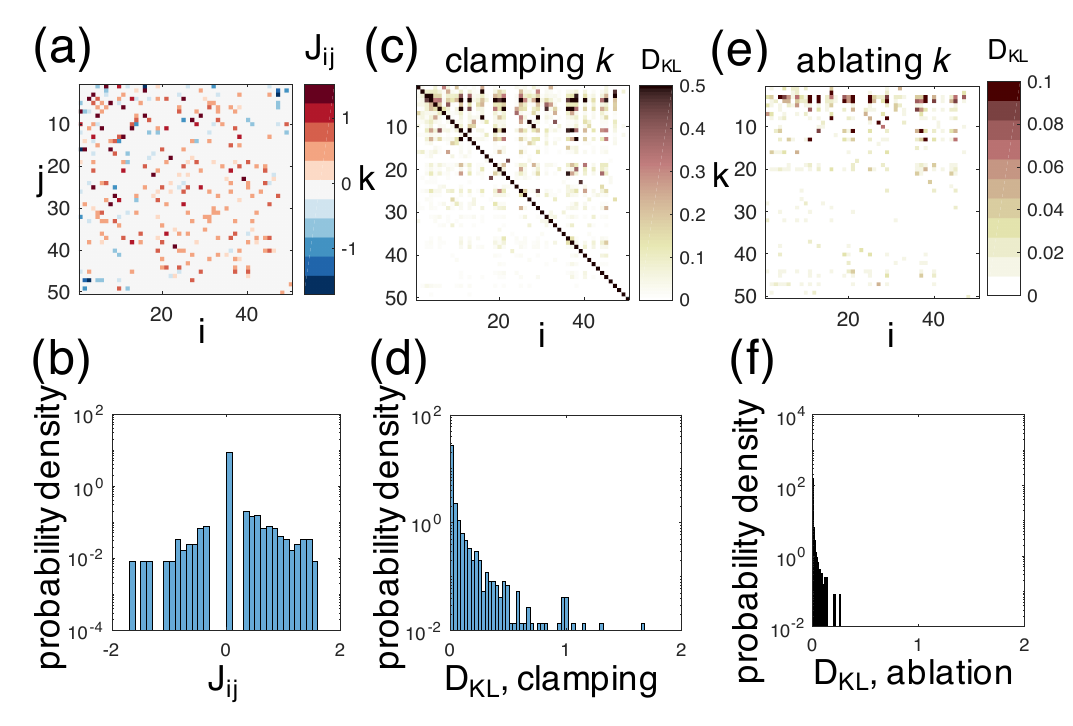}
\caption{Local perturbation of the neural network leads to global response. (a, b) For a typical group of $N = 50$ neurons, the inferred interaction matrix $J$ is sparse. Here, the neuron index $i$ and $j$ are sorted based on $m_i^\text{flat}$, as in Fig.~\ref{fig:model_construction}. (c, d) When neuron $k$ is clamped to a constant voltage, the Kullback-Leibler divergence (in bits) of the marginal distribution of states for neuron $i$ is distributed throughout the network. (e, f) When neuron $k$ is ablated, the $D_{KL}$ is also distributed throughout the network, but is smaller than in response to clamping.}
\label{fig:clamp_vs_ablation}
\end{figure}

The maximum entropy model can be perturbed through both ``clamping'' and ``ablation.''  By definition, the only possible state in which we can clamp a single neuron is the all ``flat" state, $\sigma_k = p$. Following the maximum entropy model [Eq.~(\ref{eq:hamiltonian})], the probability distribution for the rest of the network becomes
\begin{equation}
\widetilde{P}_{k}(\bm\sigma) \equiv P(\sigma_1, \sigma_2, ... \sigma_{N-1} \vert \sigma_k = 3)  = \frac{1}{\widetilde{Z}_k} e^{-\widetilde{\mathcal{H}}_k(\bm\sigma)}\,\mbox{,}
\end{equation}
where the effective Hamiltonian is
\begin{equation}
\widetilde{\mathcal{H}}_k(\bm\sigma)  = -\frac{1}{2}\sum_{i\neq j\neq k} J_{ij}\delta_{\sigma_i\sigma_j} - \sum_{i\neq k} J_{ik} \delta_{\sigma_i p} - \sum_{i\neq k}\sum_{r=1}^{p-1} h_i^r \delta_{\sigma_i r}\,\mbox{.}
\end{equation}
On the other hand, ablation of neuron $k$ means the removal of neuron $k$ from the network, which leads to an effective Hamiltonian
\begin{equation}
\widehat{\mathcal{H}}_k(\bm\sigma)  = -\frac{1}{2}\sum_{i\neq j\neq k} J_{ij}\delta_{\sigma_i\sigma_j} - \sum_{i\neq k}\sum_{r=1}^{p-1} h_i^r \delta_{\sigma_i r}\,\mbox{.}
\end{equation}

We examine the effect of clamping and ablation by Monte Carlo simulation of these modified models. We focus on the response of individual neurons $i$ to perturbing neuron $k$, which is summarized by change in the magnetizations, $m_i^r \rightarrow \tilde{m}_i^r$. But since these also represent the probabilities of finding the neuron $i$ in each of the states $r= 1, ... , p$, we can measure the change as a Kullback--Leibler divergence,
\begin{equation}
D_{KL}  = \sum_{r=1}^p m_i^r \log \left({{m_i^r}\over{\tilde{m}_i^r}} \right) \, {\rm bits}.
\end{equation}
As shown in Fig.~\ref{fig:clamp_vs_ablation}, the response of the network to the local perturbation is distributed throughout the network for both clamping and ablation. However, clamping leads to much larger $D_{KL}$s, suggesting that the network is more sensitive to clamping, and perhaps robust against (limited) ablation. Interestingly, this result echoes the experimental observation that \textit{C. elegans} locomotion is easily disturbed through optogenetic manipulation of single neurons~\cite{gordus15_opto, mochi18}, while ablation of single neurons has limited effect on the worms' ability to perform different patterns of locomotion~\cite{circuit_bargmann05, piggott11_ablate, celegans_network_control}, although further experimental investigation is needed to test our hypotheses on network response.

\section{Discussion}

Soon it should be possible to record the activity of the entire nervous system of {\em C.~elegans} as it engages in reasonably natural behaviors.   As these experiments evolve, we would like to be in a position to ask question about collective phenomena in this small neural network, perhaps discovering aspects of these phenomena which are shared with larger systems, or even (one might hope) universal.  We start modestly, guided by the state of the data.

We have built maximum entropy models for groups of up to $N=50$ cells, matching the mean activity and pairwise correlations in these subnetworks.  Perhaps our most important result is that these models work, providing successful quantitative predictions for many higher order statistical structures in the network activity. This parallels what has been seen in systems where the neurons generate action potentials, but the {\em C.~elegans} network operates in a very different regime.  The success of pairwise models in this new context adds urgency to the question of when and why these models should work, and when we might expect them to fail.

Beyond the fact that the models make successful quantitative predictions, we find other similarities with analyses of vertebrate neural networks.  The probability distributions that we infer have multiple peaks, corresponding to a rough energy landscape, and the parameters of these models appear close to a critical surface.  In addition, we have shown that the inferred model is sparse, and has topological properties similar to that of the structural connectome. Nevertheless, global response is observed when the modeled network is perturbed locally, in a way similar to experimental observations.

With the next generation of experiments, we hope to extend our analysis in four ways. First, longer recording will allow construction of meaningful models for larger groups of neurons.  If coupled with higher signal--to--noise ratios, it should also be possible to make a more refined description of the continuous signals relevant to {\em C.~elegans} neurons, rather than having to compress our description down to a small number of discrete states.   Second, registration and identification of the observed neurons will make it possible to compare the anatomical connections between neurons with the pattern of interactions in our probabilistic models. Being able to identify neurons across multiple worms will also allow us to address the degree of reproducibility across individuals, and perhaps extend the effective size of data sets by averaging. Third, optogenetic tools will allow local perturbation of the neural network experimentally, which can be compared directly with the theoretical predictions in \S V.D above. Finally, improvements in experimental methods will enable constructions of maximum entropy models for freely moving worms, with which we can map the  relation between the collective behavior identified in the neuronal activity and the behavior of the animal.

\section*{Acknowledgement}

We thank F Beroz, AN Linder, L Meshulam, JP Nguyen, M Scholz, NS Wingreen, and B Xu for many helpful discussions.  Work supported in part by the National Science Foundation through the Center for the Physics of Biological Function (PHY--1734030), the Center for the Science of Information (CCF--0939370), and PHY--1607612; and by the Simons Collaboration on the Global Brain.

\section*{Author Contributions}
XC and WB performed the analyses and the simulations. FR and AML designed and carried out the experiments. All authors contributed to the manuscript preparation.

\section*{Appendix A: Perturbation methods for overfitting analysis}

To test if our maximum entropy model overfits, we partition the samples into a set of training data and a set of test data. The difference of the per-neuron log-likelihood for the training data and the test data is used as a metric of whether the model overfits: if the two values for the log-likelihood are equal within error bars, then the model generalizes well to the test data and does not overfit. Here, we outline a perturbation analysis which uses the number of independent samples and the number of parameters of the model to estimate the expectation value of this log-likelihood difference.

Consider a Boltzmann distribution parameterized by $g = g_1, g_2, \dots, g_m$ acting on observables $\phi_1, \phi_2, \dots, \phi_m$. The probability for the $N$ spins taking the value $\sigma = \sigma_1, \sigma_2, \dots, \sigma_N$ is
\begin{equation}
P(\sigma\vert g) = \frac{1}{Z(g)}\exp\left(-\sum_{i = 1}^m g_i \phi_i(\sigma)\right)\,\mbox{,}
\end{equation}
where $Z$ is the partition function. Then, the log-likelihood of a set of data with $T$ samples under the Boltzmann distribution parameterized by $g$ is
\begin{equation}
\begin{split}
L(\sigma^1, \sigma^2, \dots, \sigma^T \vert g) & = \frac{1}{T} \sum_{t = 1}^T  \log P(\sigma^t \vert g) \\
& = -\log Z(g) - \sum_{i = 1}^m g_i \left(\frac{1}{T}\sum_{t=1}^T\phi_i^t\right)
\end{split}
\end{equation}

Now, let us assume that a set of true underlying parameters, $\lbrace g^* \rbrace$, exists for the system we study, which leads to a true expectation value be $f_i^* = f_i(g^*)$.  However, we are only given finite number of observations, $\sigma^1, \sigma^2, \dots, \sigma^T$, from which we construct a maximum entropy model, i.e. infer the parameters $\lbrace \hat{g} \rbrace$ by maximizing the likelihood of the data. Our hope is that the difference between the true parameters and the inferred parameters is small, in which case we can approximate the inferred parameters using a linear approximation
\begin{eqnarray}
g_i & = &g_i^* + \delta g_i\mbox{,} \\
\text{where} \hspace{1cm} \delta g_i & \approx & \sum_j\frac{\partial g_i}{\partial f_j}\delta f_j = -\sum_j \widetilde{\chi}_{ij} \delta f_j\mbox{.}
\end{eqnarray}
Here, $\widetilde{\chi}$ is the inverse of the susceptibility matrix $\chi_{ij} = -\partial f_i/\partial g_j =  \langle \phi_i \phi_j \rangle - \langle \phi_i \rangle \langle \phi_j\rangle$; and $\delta f_j$ is the difference between empirical mean and the true mean of $\phi_j$,
\begin{equation}
\delta f_j = \frac{1}{T}\sum_{t=1}^T \phi_j(\sigma^t) - f_j^*
\end{equation}
For convenience, we will use short-hand notation $\phi_i(\sigma^t) = \phi_i^t$ to indicate the value of the observable $\phi_i$ at time $t$.

Let the number of samples in the training data be $T_1$, and the number of samples in the test data be $T_2$. For simplicity, assume that all samples are independent. We maximize the entropy of the model on only the training data to obtain parameters $\lbrace \hat{g} \rbrace$, and we would like to know how well our model generalize to the test data. Thus, we quantify the degree of overfitting by the difference of likelihood of the training data and the test data:
\begin{widetext}
\begin{equation}\label{eq:diff_likelihood}
\begin{split}
L_\text{test} - L_\text{train} & = \left[-\log Z(\hat{g}) - \sum_{i = 1}^m \hat{g}_i \left(\frac{1}{T_2}\sum_{t'=1}^{T_2}\phi_i^{t'}\right)\right] -\left[-\log Z(\hat{g}) - \sum_{i = 1}^m \hat{g}_i \left(\frac{1}{T_1}\sum_{t=1}^{T_1}\phi_i^t\right)\right]\\
& = \sum_{i = 1}^m \left(g_i^* - \sum_j \widetilde{\chi}_{ij} \left(\frac{1}{T_1}\sum_{t=1}^{T_1} \phi_j^t - f_j^*\right)\right)\left(\frac{1}{T_1}\sum_{t=1}^{T_1}\phi_i^{t}-\frac{1}{T_2}\sum_{t'=1}^{T_2}\phi_i^{t'}\right) \,\mbox{.}
\end{split}
\end{equation}
\end{widetext}
For simplicity of notation, let us write
\begin{eqnarray}
\alpha_i^{(1)} = \frac{1}{T_1}\sum_{t=1}^{T_1}\phi_i^t-f_i^* \,\mbox{,}\hspace{1cm}
\alpha_i^{(2)} = \frac{1}{T_2}\sum_{t=1}^{T_2}\phi_i^t-f_i^*\,\mbox{.}
\end{eqnarray}
By the Central Limit Theorem, $\alpha_i^{(1)}$ and $\alpha_i^{(2)}$ are Gaussian variables. Terms that appear  in the likelihood difference [Eq.~(\ref{eq:diff_likelihood})], have expectation values
\begin{equation}
\langle \alpha_i^{(1)}\rangle = 0\,\mbox{,}\hspace{1cm}
\langle \alpha_i^{((1)}\alpha_j^{(1)} \rangle  = \dfrac{1}{T_1}\chi_{ij}\,\mbox{.} 
\end{equation}
In addition, because we assume that the training data and the test data are independent, the cross-covariance between the training data and the test data is 
\begin{equation}
\langle \alpha_i^{((1)}\alpha_j^{(2)} \rangle = 0\,\mbox{.}
\end{equation}

Combining all the above expressions, we obtain the expectation value of the likelihood difference [Eq.~(\ref{eq:diff_likelihood})],
\begin{equation}
\begin{split}
\langle L_\text{test} - L_\text{train} \rangle & =\Big\langle \sum_{i = 1}^m \left(g_i^* - \sum_j \widetilde{\chi}_{ij} \alpha_{j}^{(1)} \right)\left( \alpha_i^{(1)} - \alpha_i^{(2)} \right)\Big\rangle \\
& = -\sum_{i = 1}^m \sum_{j = 1}^m \widetilde{\chi}_{ij} \langle \alpha_i^{(1)}\alpha_j^{(1)} \rangle \\
& = - \frac{1}{T_1}\sum_{i = 1}^m \sum_{j = 1}^m \widetilde{\chi}_{ij} \chi_{ij}  \\
& = -\frac{m}{T_1}
\end{split}
\end{equation}
Note that the difference of likelihood is only related to the number of parameters in our model and the number of independent samples in the training data.

Similarly, we can evaluate the variance of the likelihood difference to be
\begin{eqnarray}
\langle(L_\text{test} - L_\text{train})^2\rangle &=& \sum_{i,k}g_i^*g_k^*\chi_{ik}\left(\frac{1}{T_1}+\frac{1}{T_2}\right) \nonumber\\&&+ \frac{1}{T_1^2}(m^2+2m)+\frac{m}{T_1T_2}
\end{eqnarray}
using Wick's theorem for multivariate Gaussian variables and chain rules of partial derivatives.

In order to test whether perturbation theory can be applied to the maximum entropy model learned from the real data, we estimate the number of independent samples using $N_\text{ind. sample} \sim T/\tau$, where $T$ is the length of the experiment and $\tau$ is the correlation time. The correlation time is extracted as the decay exponent of the overlap function, defined to be 
\begin{equation}
q(\Delta t) = \Big\langle\frac{1}{N}\sum_{i=1}^N\delta_{\sigma_i(t)\sigma_i(t+\Delta t)} \Big\rangle_t\,\mbox{,}
\end{equation}
In our experiment, the correlation time is $\tau = 4 \sim 6$s. For a typical recording of 8 minutes, the number of independent samples is between 80 and 120.

In Figure~\ref{fig:overfit}, we compute the perturbation results using the number of non-zero parameters after the training and the number of independent samples estimated from the data. The prediction is within the error bar from the data, which suggests that the inferred coupling is within the perturbation regime of the true underlying coupling. Note that the plotted difference is computed for the per-neuron log-likelihood, $l_\text{test} - l_\text{train} = (L_\text{test} - L_\text{train})/N$.

\section*{Appendix B: Maximum entropy model with the pairwise correlation tensor constraint}

To fully describe the pairwise correlation between neurons with $p = 3$ states, the equal-state pairwise correlation $c_{ij} = \langle \delta_{\sigma_i\sigma_j}\rangle$ is not enough; rather, we should constrain the pairwise correlation tensor, defined as
\begin{equation}
c_{ij}^{rs} \equiv \langle \delta_{\sigma_i r}\delta_{\sigma_j s}\rangle\,\mbox{.}
\end{equation} 
Here, we constrain the pairwise correlation tensor $c_{ij}^{rs}$ together with the local magnetization $m_{i}^r \equiv \langle \delta_{\sigma_i r} \rangle$. Notice that for each pair of neurons $(i, j)$, the number of constraints are $p^2+2p = 15$, but these constraints are related through normalization requirements, $\sum_r m_i^r = 1$ and $\sum_s c_{ij}^{rs} = m_i^r$, which leads to only 7 independent variables for each pair of neurons. Because of this nonindependence, choosing which variables to constraint is a problem of gauge fixing. Here, we choose the gauge where we constrain the local magnetization $m_i^r$ for states ``rise" and ``fall", and  the pairwise correlations $c_{ij}^r \equiv c_{ij}^{rr}$; in this gauge  the parameters can be compared meaningfully  to the equal-state maximum entropy model above. The corresponding maximum entropy model has the form
\begin{equation}\label{eq:tensor_maxent}
P(\sigma) \propto \exp\left(-\frac{1}{2}\sum_{i\neq j}\sum_{r=1}^3 J_{ij}^r \delta_{\sigma_i r}\delta_{\sigma_j r} -\sum_{i}\sum_{r=1}^2 h_i^r \delta_{\sigma_i r} \right)
\end{equation}

Note that the equivalence between constraining the equal-state correlation for each state and constraining the full pairwise correlation tensor only holds for the case of $p = 3$. For $p > 3$ states, one need to choose more constraints to fix the gauge, and it is not obvious which variables to fix.

We train the maximum entropy model with tensor constraint [Eq.~(\ref{eq:tensor_maxent})] with the same procedure as the model with equal-state correlation constraint, described in the main text. The model is able to reproduce the constraints with a sparse interaction tensor $J$. However, as shown in the bottom panel of Fig.~\ref{fig:overfit}, the difference between $l_\text{train}$, the per-neuron log likelihood of the training data (randomly chosen $5/6$ of all data) and $l_\text{test}$, the per-neuron log likelihood of the test data, is greater than zero within error bars. This indicates that the maximum entropy model with tensor constraint overfits for all $N = 10,\, 20,\, \dots,\, 50$.

\section*{Appendix C: Maximum entropy model fails to predict the dynamics of the neural networks as expected}

\begin{figure}
\includegraphics[width=\linewidth]{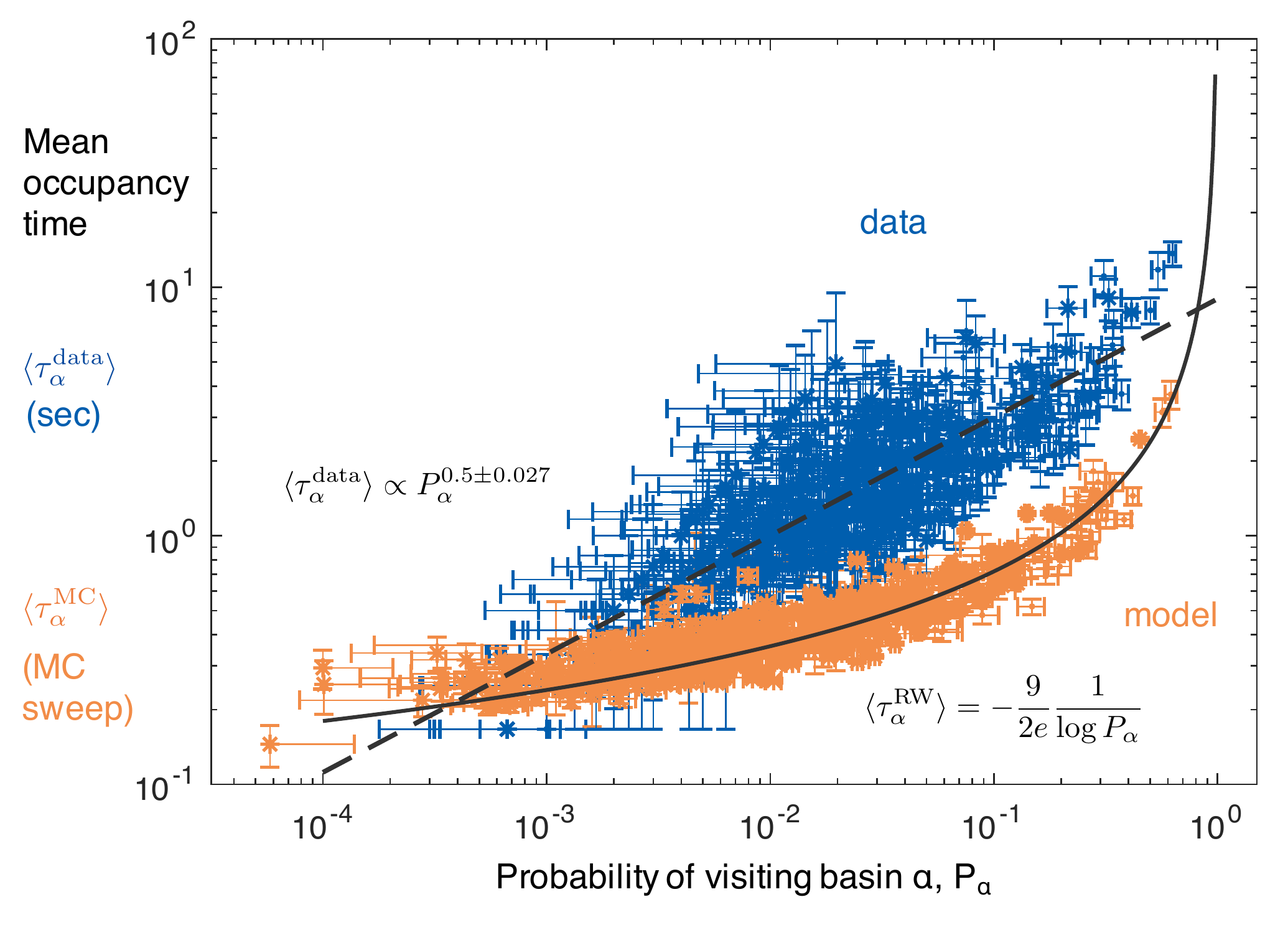}
\caption{Equilibrium dynamics of the inferred pairwise maximum entropy model fails to capture the neural dynamics of \textit{C. elegans}. The mean occupancy time of each basin of the energy landscape, $\langle\tau_\alpha\rangle$, is plotted against the fraction of time the system visits the basin, $P_\alpha$. For 10 subgroups of $N = 10$ (in dots) and $N = 20$ (in asterisks) neurons, the empirical dynamics exhibits a weak power-law relation between $\langle\tau_\alpha\rangle$ and $P_\alpha$. The striped patterns are artifacts due to finite sample size. In contrast, equilibrium dynamics extracted from a Monte Carlo simulation following detailed balance shows an inverse logarithmic relation between $\langle \tau_\alpha \rangle$ and $P_\alpha$, which can be explained by random walks on the energy landscape. Error bars of the data are extracted from random halves of the data. Error bars of the Monte Carlo simulation are calculated using correlation time and standard deviation of the observables.}
\label{fig:dynamics}
\end{figure}

By construction, the maximum entropy model is a static probability model of the observed neuronal activities. No constraint on the dynamics was imposed in building the model, and infinitely many dynamical models can generate the observed static distribution. The simplest possibility corresponds to the dynamics being like the dynamics of Monte Carlo itself, which is essentially Brownian motion on the energy landscape.  To test whether this equilibrium dynamics can capture the real neural dynamics of \textit{C. elegans}, we compare the mean occupancy time of each basin, $\langle\tau_\alpha\rangle$, calculated using the experimental data and using MCMC. The mean occupancy time is defined as the average time a trajectory spends in a basin before escaping to another basin. For equilibrium dynamics, the mean occupancy time is determined by the height of energy barriers according to the transition state theory, or by considering random walks on the energy landscape, which gives the relation $\tau \sim -p^2/2e\ln(P_\alpha)$, where $p = 3$ is the number of Potts states and $P_\alpha$ is the fraction of time the system visits basin $\alpha$. As shown in Figure~\ref{fig:dynamics}, the mean occupancy time $\langle\tau_\alpha^\text{MC}\rangle$ found in the Monte Carlo simulation can be predicted by this simple  approximation. In contrast, the empirical neural dynamics deviates from the equilibrium dynamics, as we might have  expected. The dependence between $\langle\tau_\alpha^\text{data}\rangle$ and $P_\alpha^\text{data}$ is weak; a linear fit gives $\langle\tau_\alpha^\text{data}\rangle \approx {P_\alpha^\text{data}}^{0.5\pm 0.027}$.

\bibliography{manuscript_ref.bib}

\end{document}